\documentclass[twocolumn,showpacs,preprintnumbers,amsmath,amssymb,superscriptaddress,floatfix,nofootinbib]{revtex4}

\usepackage{graphicx}
\usepackage{amsmath}
\usepackage{amsfonts}
\usepackage{amssymb}

\begin{document}

\title{The three-pion decays of the $a_1(1260)$}
\date{\today}

\author{Xu Zhang}
\affiliation{Institute of Modern Physics, Chinese Academy of
Sciences, Lanzhou 730000, China} \affiliation{University of Chinese
Academy of Sciences, Beijing 101408, China}

\author{Ju-Jun Xie}~\email{xiejujun@impcas.ac.cn}
\affiliation{Institute of Modern Physics, Chinese Academy of
Sciences, Lanzhou 730000, China} \affiliation{University of Chinese
Academy of Sciences, Beijing 101408, China}

\begin{abstract}

We investigate the decay of $a^+_1(1260) \to \pi^+ \pi^+ \pi^-$ with
the assumption that the $a_1(1260)$ is dynamically generated from
the coupled channel $\rho \pi$ and $\bar{K}K^*$ interactions. In
addition to the tree level diagrams that proceed via $a^+_1(1260)
\to \rho^0 \pi^+ \to \pi^+\pi^+ \pi^-$, we take into account also
the final state interactions of $\pi\pi \to \pi \pi$ and $K \bar{K}
\to \pi \pi$. We calculate the invariant $\pi^+ \pi^-$ mass
distribution and also the total decay width of $a^+_1(1260) \to
\pi^+ \pi^+ \pi^-$ as a function of the mass of $a_1(1260)$. The
calculated total decay width of $a_1(1260)$ is significantly
different from other model calculations and tied to the dynamical
nature of the $a_1(1260)$ resonance. The future experimental
observations could test of model calculations and would provide vary
valuable information on the relevance of the $\rho \pi$ component in
the $a_1(1260)$ wave function.

\end{abstract}

\maketitle

\section{Introduction} \label{sec:introduction}

In the naive quark model, mesons are composed of a quark-antiquark
pair. This picture works extremely well for most of the known
mesons~\cite{Patrignani:2016xqp}. However, there are a growing set
of experimental observations of resonance-like structures, which
cannot be explained by the quark-antiquark
model~\cite{Patrignani:2016xqp,Klempt:2007cp,Brambilla:2014jmp}.
Even among the seemingly well-established and understood mesons,
some of them may be more complicated than originally
thought~\cite{Baru:2003qq,Hyodo:2008xr}. One such example is the
lowest-lying axial-vector mesons. The $a_1(1260)$ is a ground state
of axial-vector resonance with quantum numbers $I^G(J^{PC}) =
1^-(1^{++})$. However, it was found that the $a_1(1260)$ could be
dynamically generated from the interactions of $K^*\bar{K}$ and
$\rho \pi$ channels and the couplings of the $a_1(1260)$ to these
channels can be also obtained at the same time~\cite{Roca:2005nm}.
Based on these results, the radiative decay of $a_1(1260)$ meson was
studied in Refs.~\cite{Roca:2006am,Nagahiro:2008cv}, where the
theoretical calculations agree with the experimental values within
uncertainties. In Ref.~\cite{Lang:2014tia} the lattice result for
the coupling constant of $a_1(1260)$ to the $\rho \pi$ channel is
also close to the value obtained in Ref.~\cite{Roca:2005nm}.
Besides, the effects of the next-to-leading order chiral potential
on the dynamically generated axial-vector mesons were studied in
Ref.~\cite{Zhou:2014ila}. It was found that the inclusion of the
higher-order kernel does not change the results obtained with the
leading-order kernel in any significant way, which gives more
supports to the dynamical picture of the $a_1(1260)$
state~\cite{Roca:2005nm,Zhou:2014ila,Lutz:2003fm}.

On the other hand, it is suggested that the $a_1(1260)$ resonance is
a candidate of the chiral partner of the $\rho$
meson~\cite{Weinberg:1967kj,Bernard:1975cd,Ecker:1988te} described
as a $q \bar{q}$ state. The nature of $a_1(1260)$ has been studied
by calculating physical observables such as the $\tau$ decay
spectrum into three
pions~\cite{GomezDumm:2003ku,Wagner:2008gz,Dumm:2009va,Nugent:2013hxa}
or the multipions decays of light vector
mesons~\cite{Achasov:2004re,Lichard:2006kw}. Recently, the
production of $a_1(1260)$ resonance in the reaction of $\pi^- p \to
a^-_1(1260) p$ within an effective Lagrangian approach was studied
in Ref.~\cite{Cheng:2016hxi} based on the results obtained in chiral
unitary approach. Furthermore, a general method was developed in
Ref.~\cite{Nagahiro:2011jn} to analyze the mixing structure of
hadrons consisting of two components of quark and hadronic
composites, and the nature of the $a_1(1260)$ was explored with the
method~\cite{Nagahiro:2011jn}, where it was found that the
$a_1(1260)$ resonance has comparable amounts of the elementary
component $q\bar{q}$ to the $\rho \pi$. In Ref.~\cite{Geng:2008ag},
the $N_c$ behavior of $a_1(1260)$ was studied using the unitarized
chiral approach, and it was found that the main component of
$a_1(1260)$ is not $q\bar{q}$. A probabilistic interpretation of the
compositeness at the pole of a resonance was been derived in
Ref.~\cite{Guo:2015daa}, where it was obtained that, for
$a_1(1260)$, the compositeness and elementariness are similar.
Furthermore, the $a_1(1260)$ can also appear as a gauge boson of the
hidden local symmetry~\cite{Bando:1987br,Kaiser:1990yf}, which is
recently reconciled with the five-dimensional gauge field of the
holographic QCD~\cite{Sakai:2004cn,Sakai:2005yt}. Yet, the nature of
the $a_1(1260)$ state is still not well understood. The only way to
understand its nature is to examine it from all possible
perspectives, both experimentally and theoretically.

On the experimental side, for the $a_1(1260)$ resonance, the
experimental width $\Gamma_{a_1(1260)}=(250-600)$ MeV assigned by
the Particle Data Group (PDG)~\cite{Patrignani:2016xqp} has large
uncertainty. While most experiments and phenomenological extractions
agree on the mass of the $a_1(1260)$ leading to a PDG value of
$M_{a_1(1260)}$ = 1230 $\pm$ 40 MeV, which is more precisely than
its width. A new COMPASS measurement in Ref.~\cite{Alekseev:2009aa}
provides a much smaller uncertainty of the width $\Gamma_{a_1(1260)}
= 367 \pm 9 ^{+28} _{-25}$ MeV and mass $M_{a_1(1260)} = 1255 \pm
6^{+7}_{-17}$ MeV. Therefore, study of the total decay width and the
decay behaviors of $a_1(1260)$ is important both on experimental and
theoretical sides, and can also provide beneficial information about
the internal structure of it.

The best knowledge about $a_1(1260)$ resonance decay channels and
branching ratio comes from hadronic $\tau$ decay
measurements~\cite{Asner:1999kj,Briere:2003fr,Coan:2004ep}, while
the $\rho \pi$ decay mode in the three-pion decays, which the
dominant decay channel of $a_1(1260)$, is the most important
one~\cite{Patrignani:2016xqp,Akhmetshin:1998df,Salvini:2004gz}. In
this work, we study the three-pion decays of the $a_1(1260)$ by
considering only the dominant $a_1(1260) \to \rho \pi$ intermediate
process and, in this calculation, we take the coupling constant of
$a_1(1260)$ to $\rho \pi$ channel in $S$-wave as that was obtained
in Ref.~\cite{Roca:2005nm}. In this respect, our calculations are
based on the dynamical picture of the $a_1(1260)$ which is a
dynamically generated state from the interactions of $\bar{K} K^*$
and $\rho \pi$ coupled channels. We calculate the energy dependence
of the partial decay width of $\Gamma_{a_1(1260) \to 3 \pi}$ as a
function of the mass of $a_1(1260)$, which could be tested by future
experiments when the precise measurements for the mass and width of
the $a_1(1260)$ resonance were done.

This article is organized as follows. In Sec.~\ref{sec:formalism},
formalism and ingredients used in the calculation are given. In
Sec.~\ref{sec:results}, the results are presented and discussed.
Finally, a short summary is given in the last section.

\section{Formalism and ingredients} \label{sec:formalism}

\begin{figure}[htbp]
\begin{center}
\includegraphics[scale=0.55]{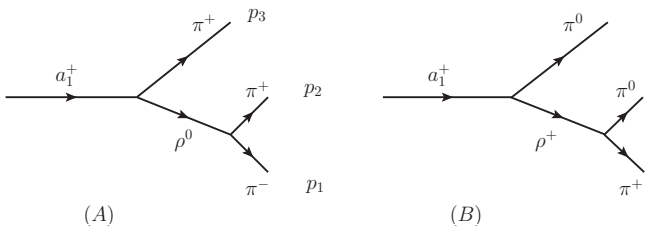}
\caption{The dominant diagrams for the decay of $a_1(1260) \to
3\pi$. (A): $a^+_1(1260) \to \pi^+ \pi^+ \pi^-$ and (B):
$a^+_1(1260) \to \pi^+ \pi^0 \pi^0$.} \label{fig:fyndig}
\end{center}
\end{figure}

We study the decay of $a_1(1260) \to 3\pi$ with the assumption that
the $a_1(1260)$ is dynamically generated from the interactions of
$\rho \pi$ and $\bar{K} K^*$ in coupled channel, thus this decay can
proceed via $a_1(1260)\to \rho \pi \to 3\pi$ as shown in
Fig.~\ref{fig:fyndig}, where we take the $a^+_1(1260) \to
\pi^+\pi^+\pi^-$ and $\pi^+ \pi^0\pi^0$ into account. It is easy to
know that the two diagrams in Fig.~\ref{fig:fyndig} give the same
contributions to the $a_1(1260) \to 3\pi$ decay. Hence, we consider
only the Fig.~\ref{fig:fyndig} in the following calculation and we
multiply by a factor two to the final result.

\subsection{Decay amplitude at tree level}

In order to evaluate the partial decay width of $a_1(1260)\to 3\pi$,
we need the decay amplitudes  of the tree level diagrams shown in
Fig. 1, where the process is described as the $a_1(1260)$ decaying
to $\rho\pi$ and then the $\rho$ decaying into $\pi\pi$. As
mentioned above, $a_1(1260)$ results as dynamically generated from
the interactions of the $\rho \pi$ and $\bar{K} K^*$ in coupled
channels. We can write the $a^+_1 \rho^0 \pi^+$ vertex as
\begin{eqnarray}
-it_1=-i\frac{g_{a_1 \rho \pi}}{\sqrt{2}}\varepsilon_{a_1}^{\mu}
\varepsilon_{\mu},
\end{eqnarray}
where $\varepsilon^\mu_{a_1}$ is the polarization vector of
$a_1(1260)$ and $\varepsilon^\mu$ the polarization vector of the
$\rho$. The $g_{a_1 \rho \pi}$ is the coupling of the $a_1(1260)$ to
the $\rho \pi$ channel and can be obtained from the residue in the
pole of the scattering amplitude in $I=1$. We take $g_{a_1 \rho \pi}
= (-3795+i2330)$ and $g_{a_1 \bar{K} K^*} = (1872 - i1468)$ MeV as
obtained in Ref.~\cite{Roca:2005nm}. We can see that the $a_1(1260)$
has large coupling to $\rho \pi$ channel comparing to the
$\bar{K}K^*$ channel.

To compute the decay amplitude, we also need the structure of the
$\rho \pi \pi$ vertices which can be evaluated by means of hidden
gauge symmetry Lagrangian describing the
vector-pseudoscalar-pseudoscalar ($VPP$)
interaction~\cite{Bando:1984ej,Bando:1987br,Meissner:1987ge,Harada:2003jx},
given by
\begin{eqnarray}
{{\cal L}_{VPP}}=-ig<V^{\mu}[P,\partial_{\mu}P]>, \label{eq:lvpp}
\end{eqnarray}
where the symbol $<>$ stands for the trace in $SU(3)$ and $g =
\frac{m_{V}}{2f}$, with $m_{V}=m_{\rho}$ and $f=93$ MeV the pion
decay constant. The matrices $P$ and $V$ contain the nonet of the
pseudoscalar mesons and the one of the vectors respectively.

From the Lagrangian of Eq.~\eqref{eq:lvpp}, the vertex of $\rho^0
\pi^+ \pi^-$ can be written as~\footnote{Note that $\sqrt{2} g$ from
the local hidden gauge approach is $5.89$, while the equivalent
quantity $g_{\rho \pi \pi}$ used in Ref.~\cite{Xie:2008ts} is
$6.05$. They differ in $2.5\%$.}
\begin{eqnarray}
{-it_2}=-ig\sqrt{2}(p_1-p_2)_{\mu}\varepsilon^{\mu},
\end{eqnarray}
where $p_1$ and $p_2$ are the momenta of $\pi^-$ and $\pi^+$ mesons,
respectively.

We can now straightforwardly construct the decay amplitude for
$a^+_1(1260) \to \pi^+ \pi^+ \pi^-$ decay corresponding to the tree
diagram shown in Fig.~\ref{fig:fyndig} (A):
\begin{eqnarray}
t_{\text{tree}} &= & - g_{a_1 \rho\pi} g \left (\frac{F_{\rho \pi\pi}(q^2_1)(p_1-p_2)_{\mu}}{q_1^2-m_{\rho}^2+im_{\rho}\Gamma_{\rho}(q^2_1)} \right .\nonumber \\
&& \left . + \frac{F_{\rho
\pi\pi}(q^2_2)(p_1-p_3)_{\mu}}{q_2^2-m_{\rho}^2+im_{\rho}\Gamma_{\rho}(q^2_2)}
\right ) \varepsilon_{a_1}^{\mu}, \\
&=&  g_{a_1 \rho\pi} g \left (\frac{F_{\rho \pi\pi}(q^2_1)(\vec{p}_1-\vec{p}_2)}{q_1^2-m_{\rho}^2+im_{\rho}\Gamma_{\rho}(q^2_1)} \right .\nonumber \\
&& \left . + \frac{F_{\rho
\pi\pi}(q^2_2)(\vec{p}_1-\vec{p}_3)}{q_2^2-m_{\rho}^2+im_{\rho}\Gamma_{\rho}(q^2_2)}
\right ) \cdot \vec{\varepsilon}_{a_1},    \label{eq:t}
\end{eqnarray}
where the two terms stand for the contributions with the $\rho^0$ in
the $\pi^-_1 \pi^+_2$ and in the $\pi^-_1 \pi^+_3$ subsystem, and
$q_1 = p_1 + p_2$ and $q_2 = p_1 + p_3$.

We take the energy dependent decay width of $\Gamma_\rho$. Because
the dominant decay channel of $\rho$ is $\pi \pi$, we take
\begin{eqnarray}
\Gamma_{\rho}(M^2_{\rm inv})=\Gamma_{{\rm on}}\left (\frac{q_{{\rm
off}}}{q_{{\rm on}}}\right )^3 \frac{m_{\rho}}{M_{\rm inv}},
\label{eq:gamrrhopipi}
\end{eqnarray}
with $\Gamma_{\rm on} = 149.1$ MeV, and
\begin{eqnarray}
q_{\rm on} &=& \frac{\sqrt{m^2_\rho - 4m^2_\pi}}{2}, \\
q_{\rm off} &= & \frac{\sqrt{M^2_{\rm inv} - 4m^2_\pi}}{2},
\end{eqnarray}
with $M^2_{\rm inv} = M^2_{12} = q^2_1$ or $q^2_2$ the invariant
mass square of the $\pi^+ \pi^-$ system corresponding the two terms
shown in Eq.~\eqref{eq:t}. We take $m_\rho = 775.26$ MeV in this
work.

It is worthy to mention that the parametrization of the width of the
$\rho$ meson shown in Eq.~\eqref{eq:gamrrhopipi} is common and it is
meant to take into account the phase space of each decay mode as a
function of the
energy~\cite{Chiang:1990ft,Xie:2007qt,Hanhart:2010wh}. In the
present work we take explicitly the phase space for the $P$-wave
decay of the $\rho$ into two pions.

Besides, in Eq.~\eqref{eq:t}, $F_{\rho\pi\pi}$ is the form factor of
$\rho^0$. In our present calculation we adopt the following form as
used in previous
works~\cite{Tsushima:2000hs,Gasparyan:2003fp,Xie:2007vs,Xie:2007qt,Xie:2015wja}
\begin{eqnarray}
F_{\rho\pi\pi}(M^2_{\rm
inv})=\frac{\Lambda_{\rho}^4}{\Lambda_{\rho}^4+(M^2_{\rm
inv}-m_{\rho}^2)^2},
\end{eqnarray}
where $\Lambda_{\rho}$ is the cutoff parameter of $\rho^0$.

\subsection{Decay amplitude for the triangular loop}

\begin{figure}[htbp]
\begin{center}
\includegraphics[scale=0.5]{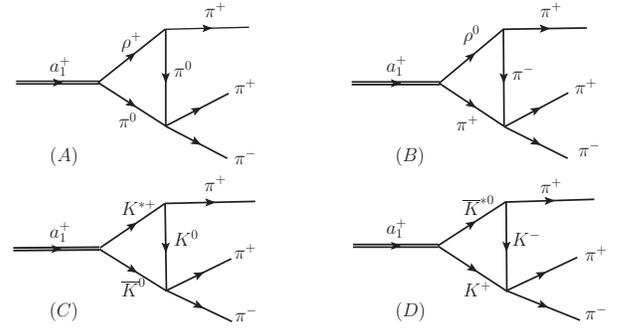}
\caption{Triangular loop contributions to the $a^+_1(1260) \to \pi^+
\pi^- \pi^+$ decay.} \label{fig:fsi}
\end{center}
\end{figure}

In addition to the tree level diagrams shown in
Fig.~\ref{fig:fyndig}, we study also the contributions of $\pi\pi
\to \pi \pi$ and $K\bar{K} \to \pi \pi$ final state interaction
(FSI). For this purpose, we use the triangular mechanism contained
in the diagrams shown in Fig.~\ref{fig:fsi}, consisting in the
rescatering of the $\pi\pi$ and $K\bar{K}$ pairs. The rescattering
of $\pi \pi$ and $K\bar{K}$ in coupled channels dynamically
generates the $f_0(500)$ and $f_0(980)$ resonances.

We can write explicitly the decay amplitudes for the triangular
diagrams shown in Fig.~\ref{fig:fsi} as (see also
Ref.~\cite{Aceti:2016yeb}, where more details can be found)
\begin{eqnarray}
t_{\rm{FSI}}^{A+B} &=& g_{a_1\rho \pi}g \left [ (2I_1+I_2)t_{\pi \pi \to \pi \pi} \vec{p}_2\cdot \vec{\varepsilon}_{a_1} \right . \nonumber\\
&& \left . + (2I_1+I_2) t_{\pi \pi \to \pi \pi } \vec{p}_3 \cdot \vec{\varepsilon}_{a_1} \right ], \label{eq:tfsiAB}\\
t_{\rm{FSI}}^{C+D} &=& -\frac{g_{a_1\bar{K} K^*} g}{\sqrt{2}} \left [ (2I'_1+I'_2)t_{K \bar{K} \to \pi \pi} \vec{p}_2\cdot \vec{\varepsilon}_{a_1} \right . \nonumber\\
&& \left . + (2I'_1+I'_2) t_{K \bar{K} \to \pi \pi } \vec{p}_3 \cdot
\vec{\varepsilon}_{a_1} \right ], \label{eq:tfsiCD}
\end{eqnarray}
with
\begin{eqnarray}
t_{\pi\pi \to \pi \pi} &=& \sqrt{2}t_{\pi^0 \pi^0 \to \pi^+ \pi^-} +
t_{\pi^+ \pi^- \to \pi^+ \pi^-}, \\
t_{K \bar{K} \to \pi \pi} &=& t_{K^0 \bar{K}^0 \to \pi^+ \pi^-} +
t_{K^+ K^- \to \pi^+ \pi^-},
\end{eqnarray}
where $t_{\pi^0 \pi^0 \to \pi^+ \pi^-}$, $t_{\pi^+ \pi^- \to \pi^+
\pi^-}$, $t_{K^0 \bar{K}^0 \to \pi^+ \pi^-}$, and $t_{K^+ K^- \to
\pi^+ \pi^-}$ are the meson-meson scattering amplitudes obtained in
the chiral unitary approach in Ref.~\cite{Oller:1997ti}, which
depend on the invariant mass of $\pi^+\pi^-$. The $t_{\pi \pi \to
\pi \pi}$ and $t_{K\bar{K} \to \pi \pi}$ in the first and second
terms in Eqs.~\eqref{eq:tfsiAB} and \eqref{eq:tfsiCD} depend on
$q^2_{2}$ and $q^2_{1}$, respectively. In addition, in
Eq.~\eqref{eq:tfsiAB} the quantities $I_1$ and $I_2$ are given by
\begin{widetext}
\begin{eqnarray}
I_1 &=& - \int \frac{d^3q}{(2\pi)^3} \frac{1}{8 \omega(q)
\omega^{\prime}(q) \omega^*(q)} \frac{1}{k^0 - \omega^{\prime}(q) -
\omega^{*}(q) + i \epsilon} \frac{1}{P^0 - \omega^*(q) - \omega(q) + i \epsilon} \nonumber \\
&& \times \frac{2P^0 \omega(q) + 2k^0 \omega^{\prime}(q) -
2(\omega(q) + \omega^{\prime}(q))(\omega(q)+\omega^{\prime}(q) +
\omega^{*}(q))}{(P^0 -
\omega(q)-\omega^{\prime}(q)-k^0+i\epsilon)(P^0+\omega(q)+\omega^{\prime}(q)-k^0-i\epsilon)}  ,\label{Eq:loopintegral1}  \\
I_2 &=& -\int \frac{d^3q}{(2\pi)^3}
\frac{\vec{k}\cdot\vec{q}/|\vec{k}|^2}{8 \omega(q)
\omega^{\prime}(q) \omega^*(q)}
\frac{1}{k^0-\omega^{\prime}(q)-\omega^{*}(q) + i \epsilon}\,\frac{1}{P^0 - \omega^*(q) - \omega(q) + i \epsilon} \nonumber \\
&& \times \frac{2P^0 \omega(q) + 2k^0 \omega^{\prime}(q)-2(\omega(q)
+ \omega^{\prime}(q))(\omega(q) + \omega^{\prime}(q) +
\omega^{*}(q))}{(P^0-\omega(q) - \omega^{\prime}(q) -k^0 + i
\epsilon)(P^0 + \omega(q) + \omega^{\prime}(q) -k^0 -i \epsilon)}
,\label{Eq:loopintegral2}
\end{eqnarray}
\end{widetext}
with $k = p_2$ for the first term and $k=p_3$ for the second term in
Eq.~\eqref{eq:tfsiAB}. While $\omega(q)  = \sqrt{\vec{q}^{~2} +
m^2_{\pi}}$, $\omega'(q) = \sqrt{(\vec q + \vec k)^2 + m^2_{\pi}}$,
and $\omega^*(q) = \sqrt{\vec{q}^{~2} + m^2_{\rho}}$ are the
energies of the $\pi^0$ ($\pi^+$) and $\pi^0$ ($\pi^-$), and $\rho$
meson in the triangular loop, respectively. A more detailed
derivation can be found in Refs.~\cite{Aceti:2015zva,Aceti:2015pma}.
Furthermore, $I'_1$ and $I'_2$ can easily be obtained just applying
the substitution to $I_1$ and $I_2$ with $m_\pi \to m_K$ and $m_\rho
\to m_{K^*}$.

It is worth mentioning that after performing the integrations, the
$I_1$ and $I_2$ integrals in the above equations depend only on the
modulus of the momentum of one of the outgoing $\pi^+$, which can be
easily related to the invariant mass of the $\pi^+ \pi^-$ system via
$M^2_{\pi^+ \pi^-} = M^2_{a_1} + m^2_{\pi} -
2M_{a_1}\sqrt{|\vec{p}_{2}|^2 + m^2_{\pi}}$ and $M^2_{\pi^+ \pi^-} =
M^2_{a_1} + m^2_{\pi} - 2M_{a_1}\sqrt{|\vec{p}_{3}|^2 + m^2_{\pi}}$
for the first and second terms in Eqs.~\eqref{eq:tfsiAB} and
\eqref{eq:tfsiCD}, respectively. The $d^3q$ integrations are done
with a cutoff $q_{\rm max} = 630$ MeV.

\section{Numerical results and discussion} \label{sec:results}

With the decay amplitudes obtained above, we can easily get the
total decay width of $a_1(1260) \to 3\pi$ which is
\begin{eqnarray}
d\Gamma=\frac{1}{192 \pi^3 M^2_{a_1}} {\sum}\vert t \vert^2  p_1^*
 p_3 dM_{12} d{\rm cos}\theta^* , \label{eq:dgamma}
\end{eqnarray}
where $t = t_{\rm tree} + t^{A+B}_{\rm FSI} + t^{C+D}_{\rm FSI}$ is
the total decay amplitude for the decay of $a^+_1(1260) \to \pi^+
\pi^+ \pi^-$. The $p_3$ and $p_1^*$ are the three-momenta of the
outgoing $\pi^+_3$ ($\pi^+_2$) meson in the $a^+_1(1260)$ rest frame
and the outgoing $\pi^-$ meson in the center of mass frame of the
final $\pi^-_1 \pi^+_2$ ($\pi^-_1 \pi^+_3$) system, respectively.
They are given by
\begin{eqnarray}
p_3 &=& \frac{\lambda^{1/2}(M^2_{a_1},M^2_{12},m^2_{\pi})}{2M_{a_1}}, \\
p^*_1 &=& \frac{\lambda^{1/2}(M^2_{12},m^2_\pi,m^2_\pi)}{2M_{12}},
\end{eqnarray}
where $\lambda(x,y,z)$ is the K\"ahlen or triangle function. We take
$m_{\pi} = 139.57$ MeV in this calculation.

For $\sum |t|^2$, the sum over polarizations can be easily done
thanks to
\begin{eqnarray}
\sum \varepsilon_{a_1}^{\mu} \varepsilon^{\nu *}_{a_1} &=& -g^{\mu
\nu} + \frac{q^{\mu}q^{\nu}}{M^2_{a_1}},
\end{eqnarray}
with $q$ the four-momentum of the $a_1(1260)$. Here we give
explicitly the results for the tree diagrams, as an example,
\begin{eqnarray}
&& \sum |t_{\rm tree}|^2 = g^2_{a_1 \rho \pi} g^2 \left( \frac{(E_1 - E_2)^2 - (p_1 - p_2)^2}{D_1} \right . \nonumber \\
&& \left . + \frac{(E_3 - E_2)^2 - (p_3-p_2)^2}{D_2} + \right . \nonumber \\
&& \left . \!\!\!\!\! \frac{(E_1-E_2)(E_3-E_2) - (p_1 -
p_2)\cdot(p_3-p_2)}{D_3} \right), \label{amplitudesquare}
\end{eqnarray}
with
\begin{eqnarray}
D_1 &=& (q_1^2 - m^2_{\rho})^2 + [m_{\rho} \Gamma_{\rho}(q_1^2)]^2, \\
D_2 &=& (q_2^2 - m^2_{\rho})^2 +  [m_{\rho} \Gamma_{\rho}(q_2^2)]^2, \\
D_3 &=& \frac{1}{2} \frac{D_1 D_2}{(q^2_1 - m^2_\rho)(q^2_2 -
m^2_\rho) + m^2_\rho  \Gamma_{\rho}(q_1^2) \Gamma_{\rho}(q_2^2)},
\end{eqnarray}
and
\begin{eqnarray}
&& E_3 = \frac{M^2_{a_1} + m^2_{\pi} -
M^2_{12}}{2M_{a_1}}, \\
&& E_1 = \frac{M_{a_1}-E_3}{2} - \frac{p_3p^*_1}{M_{12}}{\rm cos}\theta^* , \\
&& E_2 = \frac{M_{a_1}-E_3}{2} + \frac{p_3p^*_1}{M_{12}}{\rm cos}\theta^* , \\
&&(p_1 - p_2)^2 = 4m^2_\pi - M^2_{12}, \\
&&(p_3 - p_2)^2 = 3m^2_\pi -M_{a_1}E_3 - \frac{2 M_{a_1} p_3 p^*_1{\rm cos}\theta^*}{M_{12}}, \\
&&(p_1 - p_2) \cdot (p_3 - p_2)  =  \frac{M^2_{a_1} + m^2_{\pi}}{2}
- 2 M_{a_1}E_2.
\end{eqnarray}

The range of $M_{12}$ is
\begin{eqnarray}
M^{\rm max}_{12} & = & M_{a_1} - m_{\pi},
\nonumber \\
M^{\rm min}_{12} &=& 2 m_{\pi} . \nonumber
\end{eqnarray}

\begin{figure}[htbp]
\begin{center}
\includegraphics[scale=0.45]{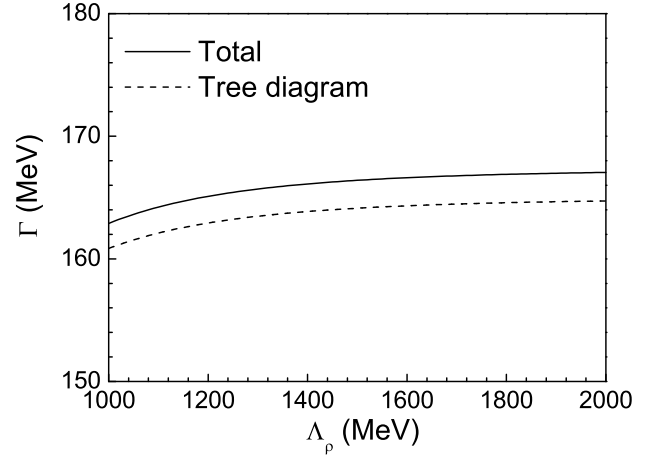}
\caption{The total decay width of $a_1(1260) \to 3\pi$ as a function
of the cutoff parameter $\Lambda_\rho$.} \label{fig:gamr-cutoff}
\end{center}
\end{figure}

With all the ingredients obtained above, one can easily get the
total decay width of $a_1(1260) \to 3 \pi$ by performing the
integration of $M_{12}$ and ${\rm cos}\theta^*$. The results for
$\Gamma$ as a function of $\Lambda_\rho$ is shown in
Fig.~\ref{fig:gamr-cutoff} with $M_{a_1} = 1230$ MeV. From
Fig.~\ref{fig:gamr-cutoff} one can see that the results for $\Gamma$
are not sensitive to the value of $\Lambda_\rho$, therefore, we fix
$\Lambda_\rho = 1500$ MeV in the next calculations.

However, since $a_1(1260)$ has large total decay width which should
be taken into account. For this purpose we replace the $d\Gamma$ in
Eq.~\eqref{eq:dgamma} by $d\tilde{\Gamma}$:
\begin{eqnarray}
d\tilde{\Gamma} = \int^{(M_{a_1} + 2\Gamma_{a_1})^2}_{(M_{a_1} -
2\Gamma_{a_1})^2} d\Gamma dm^2 S(m^2),
\end{eqnarray}
where the spectral function $S(m^2)$ is defined as
\begin{eqnarray}
S(m^2) = -\frac{1}{\pi} {\rm Im}\left( \frac{1}{m^2 - M^2_{a_1} + i
M_{a_1}\Gamma_{a_1}} \right).
\end{eqnarray}

\begin{figure}[htbp]
\begin{center}
\includegraphics[scale=0.45]{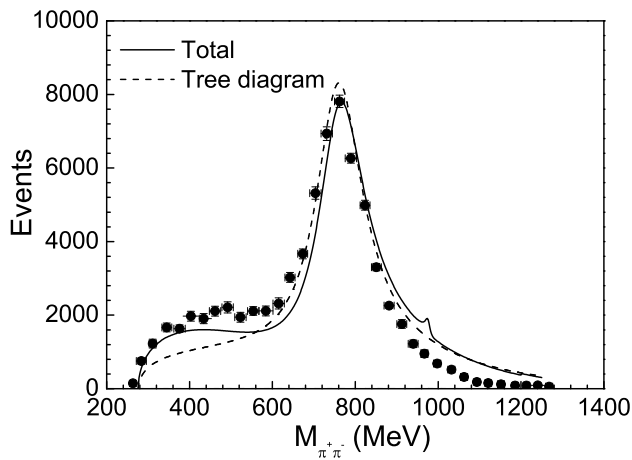}
\caption{The $\pi^+ \pi^-$ invariant mass distribution for
$a_1(1260) \to 3\pi$ as a function of the invariant mass of the
$\pi^+ \pi^-$ system. The experimental data are taken from
Ref.~\cite{Albrecht:1992ka}.} \label{fig:dgdm}
\end{center}
\end{figure}

In Fig.~\ref{fig:dgdm}, we show the numerical results for $\pi^+
\pi^-$ invariant mass distributions. We compare also our theoretical
calculations with the experimental results of
Ref.~\cite{Albrecht:1992ka} measured in the decay of $\tau \to \pi^-
\pi^- \pi^+ \nu_{\tau}$. In Fig.~\ref{fig:dgdm} we see that the tree
level alone can describe well the experimental data around the
$\rho$ peak. This is attributed to the effect of the $\rho^0$ off
shell propagator. The implementation of the contributions of the
triangle loop diagrams is responsible for the enhancement of the
invariant mass distribution at the lower invariant masses, where the
$f_0(500)$ resonance appears. There is also a small peak around the
$K\bar{K}$ mass threshold, where the $f_0(980)$ resonance appears.

The numerical results in Fig.~\ref{fig:dgdm} show how the most
drastic change in the line shape of the the invariant $\pi^+ \pi^-$
mass distribution is caused by the tree diagram alone in
Fig.~\ref{fig:fyndig} and, as mentioned before, this is tied to the
$\rho^0$ contribution, which appears at tree level because of the
large coupling of $a_1(1260)$ to $\rho \pi$ channel obtained in the
chiral unitary approach~\cite{Roca:2005nm}.

\begin{figure}[htbp]
\begin{center}
\includegraphics[scale=0.45]{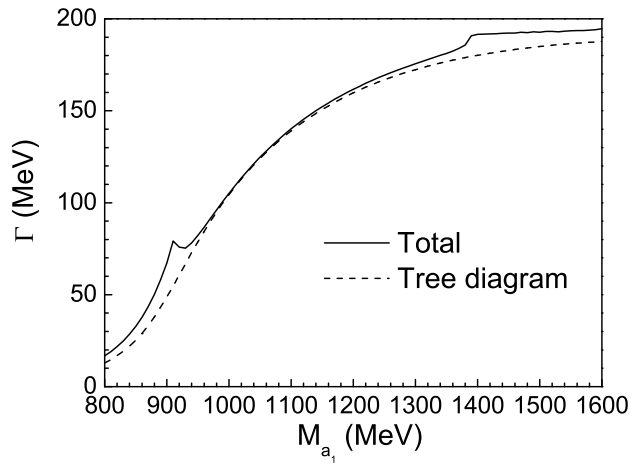}
\caption{The total decay width of $a_1(1260) \to 3\pi$ as a function
of $M_{a_1}$.} \label{fig:gamr-mass}
\end{center}
\end{figure}

Next, we calculate the total decay width of $a^+_1(1260) \to \pi^+
\pi^+ \pi^-$ as a function of the mass of $a_1(1260)$. The numerical
result is shown in Fig.~\ref{fig:gamr-mass}. The width rises rapidly
with increasing $M_{a_1}$ in the mass range $M_{a_1} < 1300$ MeV,
while it goes to flat when $M_{a_1} > 1400$ MeV. Besides, we get
$\Gamma = 166$ MeV at $M_{a_1} = 1230$ MeV. There is still no
precise measurement about the $a^+_1(1260) \to \pi^+ \pi^+ \pi^-$
decay, we cannot compare our result with experiment. Note that the
width $\Gamma_{a_1} \equiv \Gamma_{a^0_1 \to \pi^+ \pi^- \pi^0}$ was
studied in Ref.~\cite{Achasov:2004re}, and $\Gamma_{a_1} = 860$ MeV
was obtained at $M_{a_1} = 1230$ MeV. One can see that the
theoretical result in Ref.~\cite{Achasov:2004re} is much different
with us. On the other hand, there are two peaks in the solid curve
in Fig.~\ref{fig:gamr-mass}, which are attributed to the effect of
the $\pi \pi \to \pi\pi$ and $K\bar{K} \to \pi\pi$ final state
interactions. We hope that the future experiments could test the
model calculations.

So far we have assumed that the $a_1(1260)$ resonance is fully made
from $\rho \pi $ and $\bar{K} K^*$ interaction. The pole position
$(M^{\rm pole}_{a_1} - i\Gamma^{\rm pole}_{a_1}/2)$ is identified
from the zero of the denominator of the scattering amplitudes in the
complex plane, and the effective couplings $g_{a_1 \rho \pi}$ and
$g_{a_1 \bar{K}K^*}$ are calculated from the residues of the
scattering amplitudes at the complex pole. We know that the
$a_1(1260)$ Breit-Wigner parameters, $M_{a_1}$ and $\Gamma_{a_1}$,
deviate from its pole parameters by a large amount and are reaction
dependent~\cite{Patrignani:2016xqp}. On the other hand, we have no
information on how the effective couplings obtained at the pole
position change with varying $M_{a_1}$, and therefore, we cannot
include the uncertainties of these effective couplings without
making further assumptions. Besides, there are hints that the
$a_1(1260)$ resonance could have also other components as mention
above, thus, there should be also contribution from $a_1(1260) \to
f_0(500) \pi \to 3\pi$~\cite{Patrignani:2016xqp} in the tree level.
However, the information about this contribution is very scarce. We
will leave such studies to a future work.

\section{summary} \label{sec:summary}

In this work, we evaluate the partial decay width of the
$a^+_1(1260) \to \pi^+ \pi^+ \pi^-$ with the assumption that the
$a_1(1260)$ is dynamically generated from the coupled channel $\rho
\pi$ and $\bar{K}K^*$ interactions. The dominant tree level diagrams
that proceed via $a^+_1(1260) \to \rho^0 \pi^+ \to \pi^+\pi^+ \pi^-$
are considered. Besides, we also take into account the final state
interactions of $\pi\pi \to \pi\pi$ and $K\bar{K} \to \pi \pi$. It
is found that the contributions from $\pi\pi \to \pi\pi$ and
$K\bar{K} \to \pi \pi$ are small compare to the tree level diagram,
but they change the $\pi^+ \pi^-$ invariant mass distributions of
the $a_1(1260) \to 3\pi$ decay.

The results that we obtained for the $\pi^+ \pi^-$ invariant mass
distributions are in a fair agreement with the experimental
measurements for the $\tau \to \pi^- \pi^- \pi^+ \nu_{\tau}$ decay.
This provides new support for the molecular picture of $a_1(1260)$.
Furthermore, we calculate also the total decay width as a function
of the mass of $a_1(1260)$, it is found that our result is different
with other model calculations. Thus, we hope that the further
experimental observations of the $\pi^+ \pi^-$ and $\pi^+ \pi^+
\pi^-$ mass distributions would then test these model calculations
and provide vary valuable information on the relevance of the $\rho
\pi$ component in the $a_1(1260)$ wave function.

\section*{Acknowledgments}

This work is partly supported by the National Natural Science
Foundation of China under Grant Nos. 11475227 and 11735003. It is
also supported by the Youth Innovation Promotion Association CAS
(No. 2016367).

\end{document}